\documentclass[11pt,twoside]{article}
\input{epsf.sty}
\usepackage{mathrsfs}
\usepackage{amsmath}
\usepackage{amssymb}
\usepackage{graphicx}
\usepackage{subfigure}
\usepackage{cite}
\numberwithin{equation}{section}

\newtheorem{theorem}{Theorem}
\newtheorem{proposition}{Proposition}
\newtheorem{lemma}{Lemma}
\newtheorem{corollary}{Corollary}

\title{ A new two-component integrable system with peakon solutions}
\author{Baoqiang Xia$^{1}$\footnote{E-mail address:
xiabaoqiang@126.com}, ~~Zhijun Qiao$^{2}$\footnote{Corresponding author, E-mail address:
qiao@utpa.edu}
\\
$^{1}$School of Mathematics and Statistics, Jiangsu Normal
University,\\
 Xuzhou, Jiangsu 221116, P. R. China
\\
$^2$Department of Mathematics, University of Texas-Pan American, \\Edinburg, Texas 78541, USA}

\date{}
\begin{document}
\maketitle
\begin{abstract}
A new two-component system with cubic nonlinearity and linear dispersion:
\begin{eqnarray*}
\left\{\begin{array}{l}
m_t=bu_{x}+\frac{1}{2}[m(uv-u_xv_x)]_x-\frac{1}{2}m(uv_x-u_xv),
\\ n_t=bv_{x}+\frac{1}{2}[ n(uv-u_xv_x)]_x+\frac{1}{2} n(uv_x-u_xv),
\\m=u-u_{xx},~~ n=v-v_{xx},
\end{array}\right. %\label{eq}
\end{eqnarray*}
where $b$ is an arbitrary real constant, is proposed in this paper. This system is shown integrable with its Lax pair, bi-Hamiltonian structure, and infinitely many conservation laws. Geometrically, this system describes a nontrivial one-parameter family of pseudo-spherical surfaces. In the case $b=0$, the peaked soliton (peakon) and multi-peakon solutions to this two-component system are derived. In particular, the two-peakon dynamical system is explicitly solved and their interactions are investigated in details. Moreover, a new integrable cubic nonlinear equation with linear dispersion
\begin{eqnarray*}
m_t=bu_{x}+\frac{1}{2}[m(|u|^2-|u_x|^2)]_x-\frac{1}{2}m(uu^\ast_x-u_xu^\ast), \quad m=u-u_{xx},
\end{eqnarray*}
is obtained by imposing the complex conjugate reduction $v=u^\ast$ to the two-component system.
The complex valued $N$-peakon solution and kink wave solution to this complex equation are also derived.

\vspace{0.5cm}


\vspace{0.5cm}

\end{abstract}
\newpage

\section{ Introduction}
In recent years, the Camassa-Holm (CH) equation \cite{CH}
\begin{eqnarray}
m_t-bu_x+2m u_x+m_xu=0, \quad m=u-u_{xx},
\label{CH}
\end{eqnarray}
where $b$ is an arbitrary constant, derived by Camassa and Holm \cite{CH} as a shallow water
wave model, has attracted much attention in the theory of soliton and integrable system.
As an integrable equation it was implied in the work of Fuchssteiner and Fokas \cite{FF1} on hereditary symmetries as a very special case.
Since the work of Camassa and Holm \cite{CH}, more diverse studies on this equation have been remarkably developed \cite{CH2}-\cite{CGI}.
The most interesting feature of the CH equation (\ref{CH}) is that it admits peaked
soliton (peakon) solutions in the case $b=0$ \cite{CH,CH2}. A peakon is a weak
solution in some Sobolev space with corner at its crest.
The stability and interaction of peakons were discussed in several references \cite{CS1}-\cite{JR}.
Moreover, in \cite {L} the author discussed the potential applications of the CH equation to tsunami dynamics.

In addition to the CH equation, other integrable models with peakon solutions have been found \cite{DP1}-\cite{NV1}.
Among these models, there are two integrable peakon equations with cubic nonlinearity, which are
\begin{eqnarray}
 m_t=bu_x+\left[ m(u^2-u^2_x)\right]_x, \quad  m=u-u_{xx},\label{cCHQ}
\end{eqnarray}
and
\begin{eqnarray}
 m_t=u^2m_x+3uu_xm, \quad  m=u-u_{xx}.\label{cCHN}
\end{eqnarray}
Equation (\ref{cCHQ}) was proposed independently by Fokas (1995) \cite{Fo}, by Fuchssteiner (1996) \cite{Fu}, by Olver and Rosenau (1996) \cite{OR}, and  by Qiao (2006) \cite{Q1, Q11}. Equation (\ref{cCHQ}) is the first cubic nonlinear integrable system possessing peakon solutions.
Recently, the peakon stability of equation (\ref{cCHQ}) with $b=0$ was worked out by Gui, Liu, Olver and Qu \cite{GLOQ}. In 2009, Novikov \cite{NV1} derived another cubic equation, which is equation (\ref{cCHN}), from the symmetry approach, and Hone and Wang \cite{HW1} gave its Lax pair, bi-Hamiltonian structure, and peakon solutions. Very recently \cite{QXL}, we derived the Lax pair, bi-Hamiltonian structure, peakons, weak kinks, kink-peakon interactional and smooth soliton solutions for the following integrable equation with both quadratic and cubic nonlinearity \cite{Fo,Fu}:
\begin{eqnarray}
 m_t=bu_x+\frac{1}{2}k_1\left[ m(u^2-u^2_x)\right]_x+\frac{1}{2}k_2(2 m u_x+ m_xu), \quad  m=u-u_{xx},\label{gCH}
\end{eqnarray}
where $b$, $k_1$, and $k_2$ are three arbitrary constants.
By some appropriate rescaling, equation (\ref{gCH}) was implied in the papers of Fokas and Fuchssteiner \cite{Fo,Fu}, where it was derived from the two-dimensional hydrodynamical equations for surface waves.
Equation (\ref{gCH}) can also be derived by applying the tri-Hamiltonian duality to the bi-Hamiltonian Gardner equation \cite{OR}.

The above shown equations are one-component integrable peakon models.
It is very interesting for us to study multi-component integrable generalizations of peakon equations.
For example, in \cite{OR,CLZ,LZ,Fa,HI,CHIP}, the authors proposed two-component generalizations of the CH equation (\ref{CH}) with $b=0$, and
in \cite{GX,SQQ} the authors presented two-component extensions of the cubic nonlinear equation (\ref{cCHN}) and equation (\ref{cCHQ}) with $b=0$.

In this paper, we propose the following two-component system with cubic nonlinearity and linear dispersion
\begin{eqnarray}
\left\{\begin{array}{l}
m_t=bu_{x}+\frac{1}{2}[m(uv-u_xv_x)]_x-\frac{1}{2}m(uv_x-u_xv),
\\ n_t=bv_{x}+\frac{1}{2}[ n(uv-u_xv_x)]_x+\frac{1}{2} n(uv_x-u_xv),
\\m=u-u_{xx},~~ n=v-v_{xx},
\end{array}\right. \label{eq}
\end{eqnarray}
where $b$ is an arbitrary real constant.
This system is reduced to the CH equation (\ref{CH}) as $v=-2$, to the cubic CH equation (\ref{cCHQ}) as $v=2u$,
and to the generalized CH equation (\ref{gCH}) as $v=k_1u+k_2$.
Moreover, by imposing the complex conjugate reduction $v=u^\ast$, equation (\ref{eq}) is reduced to a new integrable equation with cubic nonlinearity and linear dispersion
\begin{eqnarray}
m_t=bu_{x}+\frac{1}{2}[m(|u|^2-|u_x|^2)]_x-\frac{1}{2}m(uu^\ast_x-u_xu^\ast), \quad m=u-u_{xx},
\label{nlseq}
\end{eqnarray}
where the symbol $^\ast$ denotes the complex conjugate of a potential.
The above reductions of the two-component system (\ref{eq}) look very like the ones of AKNS system,
which can be reduced to the KdV equation, the mKdV equation, the Gardner equation and the nonlinear Schr\"{o}dinger equation.
We prove the integrability of system (\ref{eq}) by providing its Lax pair, bi-Hamiltonian structure, and infinitely many conservation laws.
Geometrically system (\ref{eq}) describes pseudo-spherical surfaces and thus it is also integrable in the sense of geometry.
In the case $b=0$ (dispersionless case), we show that this system admits the single-peakon of traveling wave solution as well
as multi-peakon solutions. In particular, the two-peakon dynamic system is explicitly solved and their interactions are investigated in details.
Moreover, we propose the complex valued $N$-peakon solution and kink wave solution to the cubic nonlinear complex equation (\ref{nlseq}).
To the best of our knowledge, equation (\ref{nlseq}) is the first model admitting complex peakon solution and kink solution.

The whole paper is organized as follows. In section 2, the Lax
pair, bi-Hamiltonian structure as well as infinitely many conservation laws of equation (\ref{eq}) are presented.
In section 3, the geometric integrability of equation (\ref{eq}) are studied.
In section 4, the single-peakon, multi-peakon, and two-peakon dynamics are discussed.
Section 5 shows that equation (\ref{nlseq}) admits the complex valued peakon solution and kink wave solution.
Some conclusions and open problems are described in section 6.

\section{Lax pair, bi-Hamiltonian structure and conservation laws}

Let us consider a pair of linear spectral problems
\begin{eqnarray}
\left(\begin{array}{c}\phi_{1}\\\phi_{2} \end{array}\right)_x&=&
U\left(\begin{array}{c} \phi_{1}\\\phi_{2}
\end{array}\right),\quad
U=\frac{1}{2}\left( \begin{array}{cc} -\alpha & \lambda m\\
 -\lambda  n &  \alpha \\ \end{array} \right),
\label{LPS}\\
\left(\begin{array}{c}\phi_{1}\\\phi_{2} \end{array}\right)_t&=&
V\left(\begin{array}{c} \phi_{1}\\\phi_{2}
\end{array}\right),\quad V=-\frac{1}{2}\left( \begin{array}{cc} V_{11} & V_{12} \\ V_{21} & -V_{11} \\ \end{array} \right),
\label{LPT}
\end{eqnarray}
where $\lambda$ is a spectral parameter, $m=u-u_{xx}, \ n=v-v_{xx}$, $\alpha=\sqrt{1-\lambda^2b}$, $b$ is an arbitrary constant, and
\begin{eqnarray}
\begin{split}
V_{11}=& \lambda^{-2}\alpha+\frac{\alpha}{2}(uv-u_xv_x)+\frac{1}{2}(uv_x-u_xv),
\\ V_{12}=& -\lambda^{-1}(u-\alpha u_x)-\frac{1}{2}\lambda m(uv-u_xv_x),
\\ V_{21}=&\lambda^{-1}(v+\alpha v_x)+\frac{1}{2}\lambda n(uv-u_xv_x).
\end{split}
\end{eqnarray}

The compatibility condition of (\ref{LPS}) and (\ref{LPT}) generates %is
\begin{eqnarray}
U_t-V_x+[U,V]=0.\label{cc}
\end{eqnarray}
Substituting the expressions of $U$ and $V$ given by (\ref{LPS}) and (\ref{LPT}) into (\ref{cc}), we find that (\ref{cc}) is nothing
but equation (\ref{eq}). Hence, (\ref{LPS}) and (\ref{LPT}) exactly give the Lax pair of (\ref{eq}).

Let
\begin{eqnarray}
\begin{split}
K&=\left( \begin{array}{cc} 0 &  \partial^2-1 \\
1-\partial^2  & 0 \\ \end{array} \right),
\\J&=\left( \begin{array}{cc} \partial m\partial^{-1}m\partial-m\partial^{-1}m &  \partial m\partial^{-1} n\partial+m\partial^{-1} n+2b\partial \\
\partial n\partial^{-1} m\partial+ n\partial^{-1} m+2b\partial  & \partial n\partial^{-1} n\partial- n\partial^{-1} n \\ \end{array} \right).
\label{JK}
\end{split}
\end{eqnarray}
\begin{lemma}
$J$ and $K$ are a pair of Hamiltonian operators.
\end{lemma}
{\bf Proof} \quad
It is obvious that $K$ is Hamiltonian, since it is a skew-symmetric
operator with constant-coefficient. It is easy to check $J$ is skew-symmetric. We need to prove that $J$ satisfies the Jacobi identity
\begin{eqnarray}
\langle \zeta, J'[J\eta]\theta\rangle+\langle \eta, J'[J\theta]\zeta\rangle+\langle \theta, J'[J\zeta]\eta\rangle=0,
\label{Jacb}
\end{eqnarray}
where
$\zeta=(\zeta_1,\zeta_2)^T$, $\eta=(\eta_1,\eta_2)^T$, $\theta=(\theta_1,\theta_2)^T$ are arbitrary testing functions,
and the prime-sign means the G\^{a}teaux derivative of an operator $F$ on $q$ in the direction $\sigma$ defined as \cite{Fu}
\begin{equation}
F'[\sigma]=F'(q)[\sigma]=\left.\frac{\partial}{\partial \epsilon}\right|_{\epsilon=0}F(q+\epsilon \sigma).
\label{alpha}
\end{equation}
For brevity, we introduce the notations
\begin{eqnarray}
\begin{split}
\tilde{A}&=\partial^{-1}(m\zeta_{1,x}+n\zeta_{2,x}),\quad \tilde{B}=\partial^{-1}(m\eta_{1,x}+n\eta_{2,x}),\quad \tilde{C}=\partial^{-1}(m\theta_{1,x}+n\theta_{2,x}),
\\ A&=\partial^{-1}(m\zeta_{1}-n\zeta_{2}),\quad B=\partial^{-1}(m\eta_{1}-n\eta_{2}),\quad C=\partial^{-1}(m\theta_{1}-n\theta_{2}).
\end{split}
\end{eqnarray}
By direct calculations, we arrive at
\begin{eqnarray}
\begin{split}
\langle \zeta, J'[J\eta]\theta\rangle=&\int_{-\infty}^{+\infty}[(\theta_{1,x}m_x+\theta_{2,x}n_x)\tilde{B}\tilde{A}-(\zeta_{1,x}m_x+\zeta_{2,x}n_x)\tilde{B}\tilde{C}
+\tilde{C_x}\tilde{B_x}\tilde{A}-\tilde{A_x}\tilde{B_x}\tilde{C}]dx
\\&+\int_{-\infty}^{+\infty}[(\zeta_{1,x}m-\zeta_{2,x}n)(B\tilde{C}+C\tilde{B})-(\theta_{1,x}m-\theta_{2,x}n)(B\tilde{A}+A\tilde{B})]dx
\\&+\int_{-\infty}^{+\infty}[(\zeta_{1}m+\zeta_{2}n)BC-(\theta_{1}m+\theta_{2}n)BA]dx
\\&-2b\int_{-\infty}^{+\infty}[(\zeta_{1,x}\eta_{2,x}+\zeta_{2,x}\eta_{1,x})\tilde{C}-(\eta_{2,x}\theta_{1,x}+\eta_{1,x}\theta_{2,x})\tilde{A}]dx
\\&+2b\int_{-\infty}^{+\infty}[(\zeta_{2}\eta_{1,x}-\zeta_{1}\eta_{2,x})C-(\eta_{1,x}\theta_{2}-\eta_{2,x}\theta_{1})A]dx.
\end{split}
\label{Jacb1}
\end{eqnarray}
Based on (\ref{Jacb1}), we may verify (\ref{Jacb}) directly. This completes the proof of Lemma 1.

\begin{lemma} The following relation holds
\begin{eqnarray}
\langle \zeta, J'[K\eta]\theta\rangle+\langle \eta, J'[K\theta]\zeta\rangle+\langle \theta, J'[K\zeta]\eta\rangle
+\langle \zeta, K'[J\eta]\theta\rangle+\langle \eta, K'[J\theta]\zeta\rangle+\langle \theta, K'[J\zeta]\eta\rangle=0.
\label{CHO}
\end{eqnarray}
\end{lemma}
{\bf Proof} \quad
Direct calculations yield that
\begin{eqnarray}
\begin{split}
\langle \zeta, J'[K\eta]\theta\rangle=&-\int_{-\infty}^{+\infty}[(\zeta_{2,x}\eta_1-\zeta_{1,x}\eta_2)\tilde{C}-(\theta_{2,x}\eta_1-\theta_{1,x}\eta_2)\tilde{A}]dx
\\&-\int_{-\infty}^{+\infty}[(\zeta_{1,x}\eta_{2,xx}-\zeta_{2,x}\eta_{1,xx})\tilde{C}-(\eta_{2,xx}\theta_{1,x}-\eta_{1,xx}\theta_{2,x})\tilde{A}]dx
\\&+\int_{-\infty}^{+\infty}[(\zeta_{1}\eta_{2}+\zeta_{2}\eta_{1})C-(\eta_{1}\theta_{2}+\eta_{2}\theta_{1})A]dx
\\&-\int_{-\infty}^{+\infty}[(\zeta_{1}\eta_{2,xx}+\zeta_{2}\eta_{1,xx})C-(\eta_{1,xx}\theta_{2}+\eta_{2,xx}\theta_{1})A]dx.
\end{split}
\label{CHO1}
\end{eqnarray}
Formula (\ref{CHO}) may be verified based on (\ref{CHO1}). The proof of Lemma 2 is finished.

From Lemma 1 and Lemma 2, we immediately obtain
\begin{proposition}
$J$ and $K$ are compatible Hamiltonian operators.
\end{proposition}
Furthermore, we have
\begin{proposition}
Equation (\ref{eq}) can be rewritten in the following bi-Hamiltonian form
\begin{eqnarray}
\left(m_t,~ n_t\right)^{T}=J \left(\frac{\delta H_1}{\delta m},~\frac{\delta H_1}{\delta  n}\right)^{T}=K \left(\frac{\delta H_2}{\delta m},~\frac{\delta H_2}{\delta  n}\right)^{T},\label{BH}
\end{eqnarray}
where $J$ and $K$ are given by (\ref{JK}), and
\begin{eqnarray}
\begin{split}
H_1&=\frac{1}{2}\int_{-\infty}^{+\infty}(uv+u_xv_x)dx,
\\
H_2&=\frac{1}{4}\int_{-\infty}^{+\infty}[(u^2v_x+u_x^2v_x-2uu_xv)n+2b(uv_x-u_xv)]dx.
\end{split}
\label{H}
\end{eqnarray}
\end{proposition}

Next we construct conservation laws of equation (\ref{eq}).
Let $\varphi=\frac{\phi_2}{\phi_1}$, where $\phi_1$ and $\phi_2$ are determined through equations (\ref{LPS}) and (\ref{LPT}). From (\ref{LPS}), one can easily verify that $\varphi$ satisfies the Riccati equation
\begin{eqnarray}
\varphi_x=-\frac{1}{2}\lambda m \varphi^2+ \alpha\varphi-\frac{1}{2}\lambda n.
\label{ric}
\end{eqnarray}
Equations (\ref{LPS}) and (\ref{LPT}) give rise to
\begin{eqnarray}
(\ln \phi_1)_x=-\frac{\alpha}{2}+\frac{1}{2}\lambda m\varphi,
\quad (\ln \phi_1)_t=-\frac{1}{2}\left(V_{11}+V_{12}\varphi\right),
\label{lnp}
\end{eqnarray}
which yields conservation law of equation (\ref{eq}):
\begin{eqnarray}
\rho_t=F_x,
\label{CL}
\end{eqnarray}
where
\begin{eqnarray}
\begin{split}
\rho&=m\varphi,
\\F&=\lambda^{-2}(u-\alpha u_x)\varphi-\frac{1}{2}\lambda^{-1}\left(\alpha uv-\alpha u_xv_x+uv_x-u_xv\right)+\frac{1}{2}m(uv-u_xv_x)\varphi.
\end{split}
\label{rj}
\end{eqnarray}
Usually $\rho$ and $F$ are called a conserved density and an associated flux,
respectively. In the case $b=0$, we are able to derive the explicit forms of conservation densities by expanding $\varphi$ in powers of $\lambda$ in two ways.
The first one is to expand $\varphi$ in terms of negative powers of $\lambda$ as
\begin{equation}
\varphi=\sum_{j=0}^{\infty}\varphi_j\lambda^{-j}.\label{oe1}
\end{equation}
Substituting (\ref{oe1}) into (\ref{ric}) %with $b=0$
and equating the coefficients of powers of $\lambda$, we arrive at
\begin{eqnarray}
\varphi_{0}&=\sqrt{-\frac{n}{m}}, \qquad \varphi_{1}=\frac{mn_x-m_xn-2mn}{2m^2n},
\label{w1}
\end{eqnarray}
and the recursion relation for $\varphi_{j}$:
\begin{eqnarray}
\varphi_{j+1}&=\frac{1}{m\varphi_0}\left[\varphi_j-\varphi_{j,x}-\frac{1}{2}m\sum_{i+k=j+1,~i,k\geq 1}\varphi_i\varphi_k\right],\quad j\geq 1.
\label{wj}
\end{eqnarray}
Inserting (\ref{oe1}), (\ref{w1}) and (\ref{wj}) into (\ref{rj}),
we obtain the following infinitely many conserved densities and the associated fluxes of equation (\ref{eq}): % in the case of $b=0$:
\begin{eqnarray}
\begin{split}
\rho_{0}&=\sqrt{-mn}, ~~~~ F_0=\frac{1}{2}\sqrt{-mn}(uv-u_xv_x),
\\
\rho_{1}&=\frac{mn_x-m_xn-2mn}{2mn}, ~~~~ F_1=-\frac{1}{2}(uv-u_xv_x+uv_x-u_xv)+\frac{1}{2}\rho_1(uv-u_xv_x),
\\
\rho_{j}&=m\varphi_j, ~~~~F_{j}=(u-u_x)\varphi_{j-2}+\frac{1}{2}\rho_j(uv-u_xv_x),\quad j\geq 2,
\end{split}
\label{rjj}
\end{eqnarray}
where $\varphi_j$ is given by (\ref{w1}) and (\ref{wj}).

The second expansion of $\varphi$ is in the positive powers of $\lambda$,
\begin{equation}
\varphi=\sum_{j=0}^{\infty}\varphi_j\lambda^{j}.\label{oe2}
\end{equation}
Substituting (\ref{oe2}) into (\ref{ric}) and comparing powers of $\lambda$ produce
\begin{eqnarray}
\begin{split}
\varphi_{2j}&=0, \quad j\geq 0,
\\
\varphi_{1}&=\frac{1}{2}(v+v_x) , \quad \varphi_{2j+1}-\varphi_{2j+1,x}=\frac{1}{2}m\sum_{i+k=2j,~0\leq i,k\leq 2j}\varphi_i\varphi_k,\quad j\geq 1.
\end{split}
\label{wj22}
\end{eqnarray}
By inserting (\ref{oe2}) and (\ref{wj22}) into (\ref{rj}), we arrive at
\begin{eqnarray}
\rho_{2j}=0, \quad A_{2j}=0, \quad j\geq 0,
\label{rjj2}
\end{eqnarray}
and
\begin{eqnarray}
\begin{split}
\rho_{1}&=\frac{1}{2}m(v+v_x), ~~~~ A_1=(u-u_x)\varphi_3+\frac{1}{4}m(uv-u_xv_x)(v+v_x),
\\
\rho_{2j+1}&=m\varphi_{2j+1}, ~~~~A_{2j+1}=(u-u_x)\varphi_{2j+3}+\frac{1}{2}m(uv-u_xv_x)\varphi_{2j+1},\quad j\geq 1,
\end{split}
\label{rjj3}
\end{eqnarray}
where the odd-index $\varphi_{2j+1}$ is defined by the recursion relation
\begin{eqnarray}
\varphi_{2j+1}=\frac{1}{2}(1-\partial_x)^{-1}\left(m\sum_{i+k=2j,~0\leq i,k\leq 2j}\varphi_i\varphi_k\right),\quad j\geq 1.
\label{wj23}
\end{eqnarray}
Formula (\ref{rjj2}) means that the even-index conserved densities and associated fluxes are trivial.
Formulas (\ref{rjj3}) and (\ref{wj23}) show that the nontrivial high-order odd-index conserved densities may involve in nonlocal expressions in $u$ and $v$.

\vspace*{0.2cm}
{\bf Remark 1.}
Here, we have derived two sequences of infinitely many conserved densities and the associated fluxes for equation (\ref{eq}).
The conserved densities in the sequence (\ref{rjj}) become singular when the denominators have zero points.
The conserved densities in the sequence (\ref{rjj3}) have no singularity, but they might involve in nonlocal expressions.

\section{Geometric integrability}
Based on the work of Chern and
Tenenblat \cite{CT} and the subsequent works \cite{EGR,EGR2},
a differential equation for a real valued function $u(x,t)$ is said to
describe pseudo-spherical surfaces if it is the necessary and sufficient
condition for the existence of smooth functions $f_{ij}$, $i=1,~2,~3$, $j=1,~2$,
depending on $x$, $t$, $u$ and its derivatives, such that the one-forms
$\omega_{i}=f_{i1}dx+f_{i 2}dt$ satisfy the structure equations of a surface of constant
Gaussian curvature equal to $-1$ with metric $\omega_{1}^2+\omega_{2}^2$ and connection one-form $\omega_{3}$, namely
\begin{eqnarray}
\begin{split}
d\omega_1=\omega_3\wedge\omega_2,
~~
d\omega_2=\omega_1\wedge\omega_3,
~~
d\omega_3=\omega_1\wedge\omega_2.
\end{split}
\label{se}
\end{eqnarray}

Let us consider
\begin{eqnarray}
\begin{split}
f_{11}&=-\frac{1}{2}\lambda[e^{(\alpha-\lambda)x}m-e^{(\lambda-\alpha)x}n],
\\
f_{12}&=\frac{1}{2}\lambda^{-1}[e^{(\lambda-\alpha)x}(v+\alpha v_x)-e^{(\alpha-\lambda)x}(u-\alpha u_x)]+\frac{1}{4}\lambda[e^{(\lambda-\alpha)x}n-e^{(\alpha-\lambda)x}m](uv-u_xv_x),
\\
f_{21}&=\lambda,
\\
f_{22}&=\lambda^{-2}\alpha+\frac{\alpha}{2}(uv-u_xv_x)+\frac{1}{2}(uv_x-u_xv),
\\
f_{31}&=-\frac{1}{2}\lambda[e^{(\alpha-\lambda)x}m+e^{(\lambda-\alpha)x}n],
\\
f_{32}&=-\frac{1}{2}\lambda^{-1}[e^{(\lambda-\alpha)x}(v+\alpha v_x)+e^{(\alpha-\lambda)x}(u-\alpha u_x)]-\frac{1}{4}\lambda[e^{(\lambda-\alpha)x}n+e^{(\alpha-\lambda)x}m](uv-u_xv_x),
\end{split}
\label{f}
\end{eqnarray}
and introduce the following three one-forms
\begin{eqnarray}
\begin{split}
\omega_1=f_{11}dx+f_{12}dt,
~~
\omega_2=f_{21}dx+f_{22}dt,
~~
\omega_3=f_{31}dx+f_{32}dt.
\end{split}
\label{omg}
\end{eqnarray}
Through a direct computation, we find that the structure equations (\ref{se}) hold whenever $u(x,t)$ and
$v(x,t)$ are solutions of system (\ref{eq}). Thus we have
\begin{theorem}
System (\ref{eq}) describes pseudo-spherical surfaces.
\end{theorem}

Recall that a differential equation is geometrically integrable if it describes a nontrivial one-parameter
family of pseudo-spherical surfaces. It follows that
\begin{corollary}
System (\ref{eq}) is geometrically integrable.
\end{corollary}

According to \cite{CT}-\cite{RS}, we have the following fact
\begin{proposition}
A geometrically integrable equation with associated one-forms $\omega_{i}$, $i=1, 2, 3$, is the integrability condition of a one-parameter family of $sl(2,R)$-valued linear problem
\begin{eqnarray}
d\Phi=\Omega \Phi,
\label{dv}
\end{eqnarray}
where $\Omega$ is the matrix-valued one-form
\begin{eqnarray}
\Omega=Xdx+Tdt=\frac{1}{2}\left( \begin{array}{cc} \omega_2 & \omega_1-\omega_3\\
 \omega_1+\omega_3 &  -\omega_2 \\ \end{array} \right).
 \label{Omega}
\end{eqnarray}
\end{proposition}

Therefore, the one-forms (\ref{omg}) and (\ref{dv}) yield an $sl(2,R)$-valued linear problem $\Phi_x=X\Phi$ and $\Phi_t=T\Phi$,
whose integrability condition is the two-component system (\ref{eq}). The expression (\ref{Omega}) implies that the matrices $X$ and $T$ are
\begin{eqnarray}
\begin{split}
X&=\frac{1}{2}\left( \begin{array}{cc} \lambda & \lambda e^{(\lambda-\alpha)x} n\\
 -\lambda e^{(\alpha-\lambda)x} m &  -\lambda \\ \end{array} \right),
 \\
T&=\frac{1}{2}\left( \begin{array}{cc} \lambda^{-2}\alpha+\frac{\alpha}{2}(uv-u_xv_x)+\frac{1}{2}(uv_x-u_xv) & [\lambda^{-1}(v+\alpha v_x)
+\frac{\lambda}{2}n(uv-u_xv_x)]e^{(\lambda-\alpha)x}\\
-[\lambda^{-1}(u-\alpha u_x)+\frac{\lambda}{2} m(uv-u_xv_x)]e^{(\alpha-\lambda)x} & -\lambda^{-2}\alpha-\frac{\alpha}{2}(uv-u_xv_x)-\frac{1}{2}(uv_x-u_xv)   \\ \end{array} \right).
\end{split}
 \label{XT}
\end{eqnarray}

\section{Peakon solutions to system (\ref{eq}) in the case $b=0$}
In this section, we shall derive the peakon solutions to the two-component system (\ref{eq}) with $b=0$ in two situations. The first situation is the peakon solutions with the same peakon position. The second situation is the peakon solutions with different peakon positions, which is studied by C.J. Cotter et al. \cite{CHIP} for a cross-coupled CH equation.

\subsection{Peakon solutions to the two-component system (\ref{eq}) with the same peakon position}

Let us suppose that a single peakon solution of (\ref{eq}) with $b=0$ is of the following form
\begin{eqnarray}
u=c_1e^{-\mid x-ct\mid},\quad v=c_2e^{-\mid x-ct\mid}, \label{ocp}
\end{eqnarray}
where the two constants $c_1$ and $c_2$ are to be determined.
With the help of distribution theory, we are able to write out $u_x$, $m$ and $v_x$, $n$ as follows
\begin{eqnarray}
\begin{split}
u_x&=-c_1sgn(x-ct)e^{-\mid x-ct\mid}, \quad m=2c_1\delta(x-ct),
\\
v_x&=-c_2sgn(x-ct)e^{-\mid x-ct\mid}, \quad n=2c_2\delta(x-ct).
\end{split}
\label{ocpd}
\end{eqnarray}
Substituting (\ref{ocp}) and (\ref{ocpd}) into (\ref{eq}) with $b=0$ and integrating in the distribution sense,
%against %suitable
%test functions with support around the peak,
one can readily see %we obtain
that $c_1$ and $c_2$ should satisfy
\begin{eqnarray}
c_1c_2=-3c. \label{C1}
\end{eqnarray}
In particular, for $c_1=c_2$, we recover the single peakon solution $u=\pm \sqrt{-3c}e^{-\mid x-ct\mid}$ of the cubic  CH equation (\ref{cCHQ}) with $b=0$ \cite{GLOQ,QXL}.

Let us now assume a two-peakon solution as follows:%the form
\begin{eqnarray}
u=p_1(t)e^{-\mid x-q_1(t)\mid}+p_2(t)e^{-\mid x-q_2(t)\mid}, \quad v=r_1(t)e^{-\mid x-q_1(t)\mid}+r_2(t)e^{-\mid x-q_2(t)\mid}.\label{tp}
\end{eqnarray}
In the sense of distribution, we have
\begin{eqnarray}
\begin{split}
u_x&=-p_1sgn(x-q_1)e^{-\mid x-q_1\mid}-p_2sgn(x-q_2)e^{-\mid x-q_2\mid}, \quad m=2p_1\delta(x-q_1)+2p_2\delta(x-q_2),
\\
v_x&=-r_1sgn(x-q_1)e^{-\mid x-q_1\mid}-r_2sgn(x-q_2)e^{-\mid x-q_2\mid}, \quad n=2r_1\delta(x-q_1)+2r_2\delta(x-q_2).
\end{split}
\label{tcpd}
\end{eqnarray}
Substituting (\ref{tp}) and (\ref{tcpd}) into (\ref{eq}) with $b=0$ and integrating through test functions yield the following dynamic system:
\begin{eqnarray}
\left\{\begin{array}{l}
p_{1,t}=\frac{1}{2}p_1(p_1r_2-p_2r_1) sgn(q_1-q_2)e^{ -\mid q_1-q_2\mid},\\
p_{2,t}=\frac{1}{2}p_2(p_2r_1-p_1r_2) sgn(q_2-q_1)e^{ -\mid q_2-q_1\mid},\\
q_{1,t}=-\frac{1}{3}p_1r_1-\frac{1}{2}\left(p_1r_2+p_2r_1\right)e^{ -\mid q_1-q_2\mid},\\
q_{2,t}=-\frac{1}{3}p_2r_2-\frac{1}{2}\left(p_1r_2+p_2r_1\right)e^{ -\mid q_2-q_1\mid},\\
r_{1,t}=-\frac{1}{2}r_1(p_1r_2-p_2r_1) sgn(q_1-q_2)e^{ -\mid q_1-q_2\mid},\\
r_{2,t}=-\frac{1}{2}r_2(p_2r_1-p_1r_2) sgn(q_2-q_1)e^{ -\mid q_2-q_1\mid}.\\
\end{array}\right. \label{tpode}
\end{eqnarray}
Guided by the above equations, we may conclude the following relations:
\begin{eqnarray}
p_1=Dp_2, ~~p_1r_1=A_1, ~~p_2r_2=A_2,
\label{rtp}
\end{eqnarray}
where $D$, $A_1$ and $A_2$ are three arbitrary integration constants.

\begin{figure}
\begin{minipage}[t]{0.5\linewidth}
\centering
\includegraphics[width=2.2in]{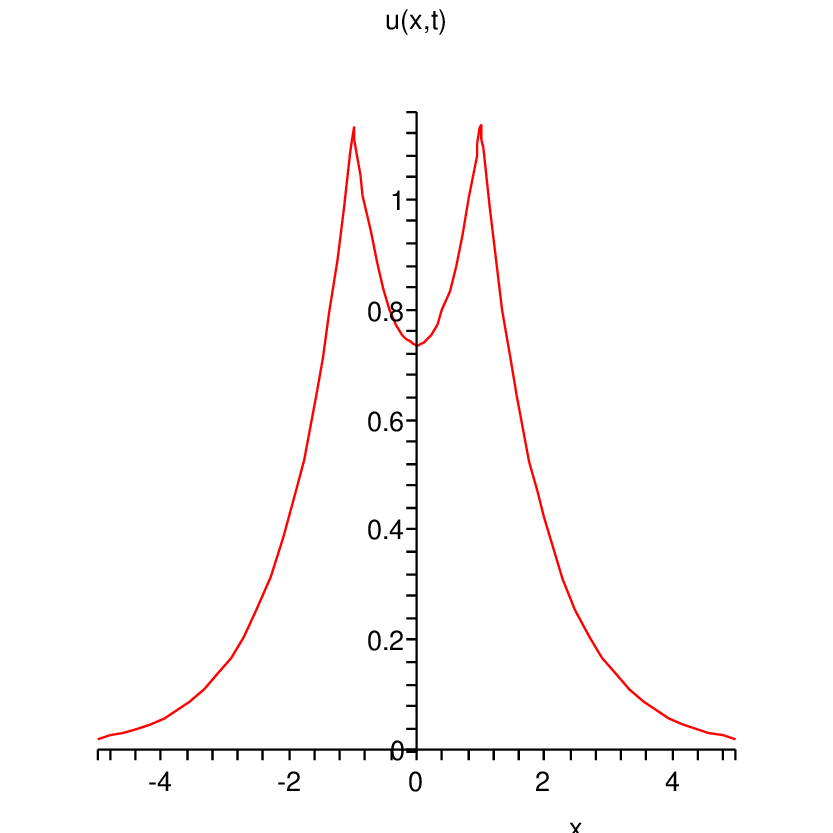}
\caption{\small{The M-shape two-peakon solution $u(x,t)$ in (\ref{tps11}) at the moment of $t=0$.}}
\label{fmp}
\end{minipage}
\hspace{2.0ex}
\begin{minipage}[t]{0.5\linewidth}
\centering
\includegraphics[width=2.2in]{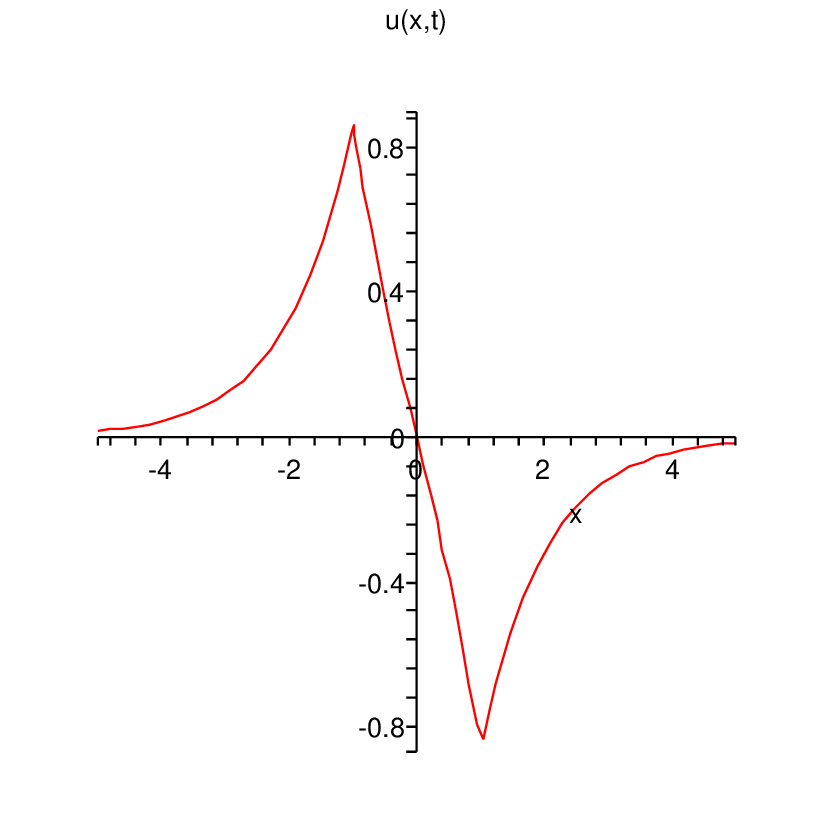}
\caption{\small{The N-shape peak-trough solution $u(x,t)$ in (\ref{tps12}) at the moment of $t=0$.}}
\label{f12}
\end{minipage}
\end{figure}

\begin{figure}
\begin{minipage}[t]{0.5\linewidth}
\centering
\includegraphics[width=2.2in]{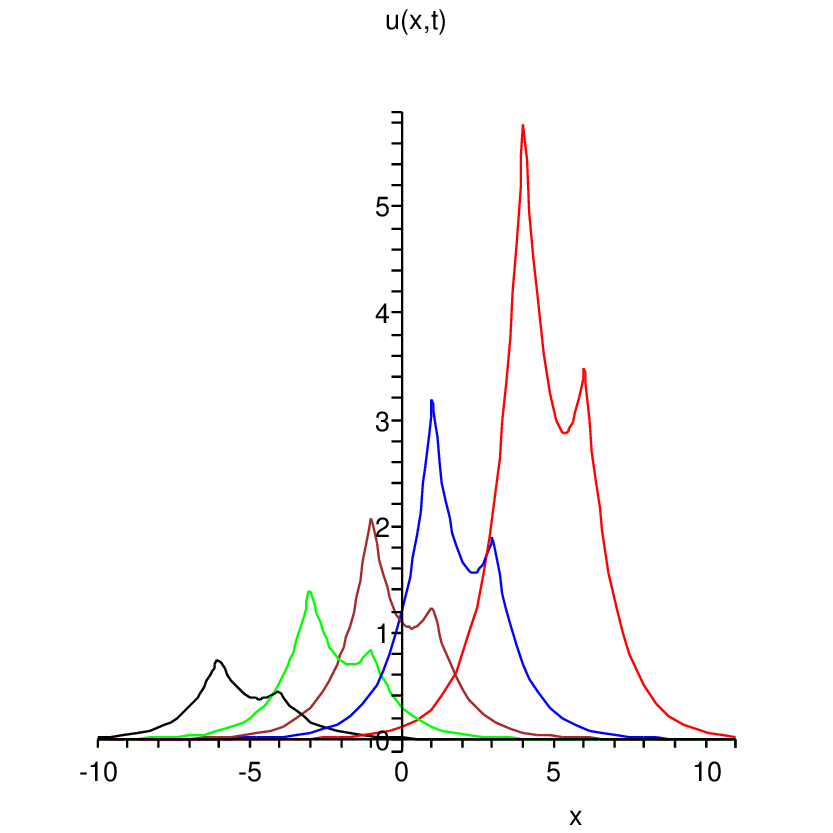}
\caption{\small{The two-peakon solution $u(x,t)$ in (\ref{tps13}). Red line: $t=-5$; Blue line: $t=-2$; Brown line: $t=0$; Green line: $t=2$; Black line: $t=5$.}}
\label{f13u}
\end{minipage}
\hspace{2.0ex}
\begin{minipage}[t]{0.5\linewidth}
\centering
\includegraphics[width=2.2in]{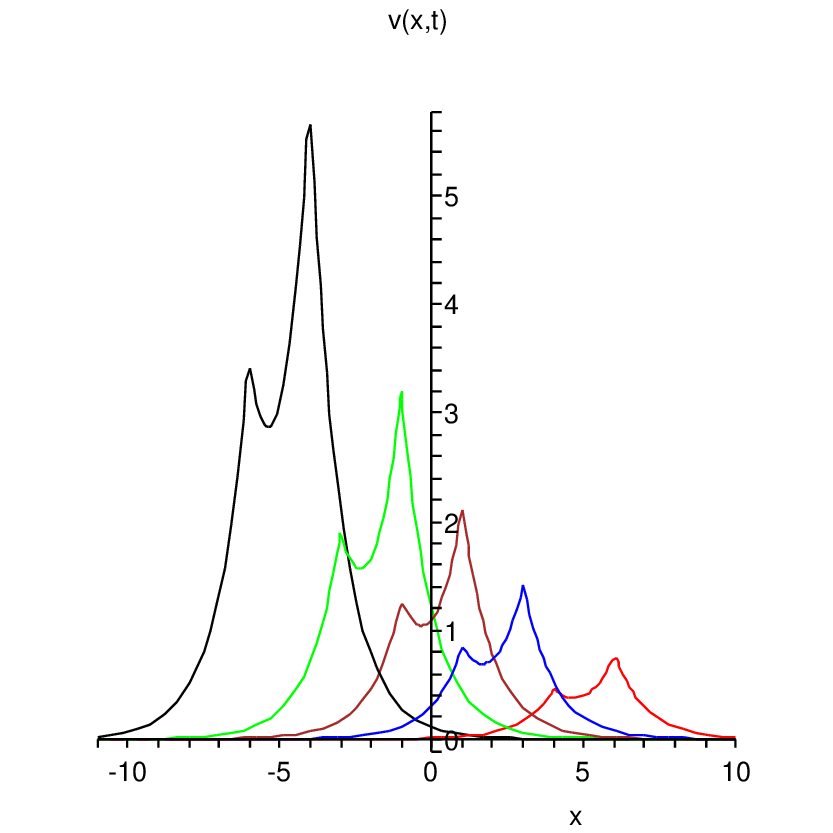}
\caption{\small{The two-peakon solution $v(x,t)$ in (\ref{tps13}). Red line: $t=-5$; Blue line: $t=-2$; Brown line: $t=0$; Green line: $t=2$; Black line: $t=5$.}}
\label{f13v}
\end{minipage}
\end{figure}

{\bf If $A_1=A_2$}, we arrive at the following solution of (\ref{tpode}):
\begin{eqnarray}
\begin{split}
p_1(t)&=Be^{\frac{1}{2D}(D^2A_1-A_1)sgn(C_1)e^{-\mid C_1\mid}t},~~
p_2(t)=\frac{p_1}{D},~~
r_1(t)=\frac{A_1}{p_1},~~
r_2(t)=\frac{A_1}{p_2},\\
q_{1}(t)&=-\left[\frac{1}{3}A_1+\frac{1}{2D}(D^2A_1+A_1)e^{-\mid C_1\mid}\right]t+\frac{1}{2}C_1,~~
q_{2}(t)=q_{1}(t)-C_1,
\end{split}
\label{tps1}
\end{eqnarray}
where $B$ and $C_1$ are two arbitrary non-zero constants. In this case, the collision between two peakons will never happen since $q_{2}(t)=q_{1}(t)-C_1$.
For example, as $ A_1=B=D=1$, $C_1=2$, (\ref{tps1}) is reduced to
\begin{eqnarray*}
\begin{split}
p_1(t)&=p_2(t)=r_1(t)=r_2(t)=1,
\\
q_{1}(t)&=-\left(\frac{1}{3}+e^{-2}\right)t+1, \qquad q_{2}(t)=-\left(\frac{1}{3}+e^{-2}\right)t-1.
\end{split}
\end{eqnarray*}
Thus, the associated solution of (\ref{eq}) with $b=0$ becomes
\begin{eqnarray}
u(x,t)=v(x,t)=e^{-\left|x+\left(\frac{1}{3}+e^{-2}\right)t-1\right|}+e^{-\left|x+\left(\frac{1}{3}+e^{-2}\right)t+1\right|}.
\label{tps11}
\end{eqnarray}
This wave has two peaks, and looks like a M-shape soliton. See Figure \ref{fmp} for this M-shape two-peakon solution.
As $ A_1=-B=-D=1$, $C_1=2$, the associated solution of (\ref{eq}) with $b=0$ becomes
\begin{eqnarray}
u(x,t)=v(x,t)=-e^{-\left|x+\left(\frac{1}{3}-e^{-2}\right)t-1\right|}+
e^{-\left|x+\left(\frac{1}{3}-e^{-2}\right)t+1\right|},
\label{tps12}
\end{eqnarray}
which has one peak and one trough and looks like N-shape soliton solution. See Figure \ref{f12} for this N-shape two-peakon solution.
As $B=2D=1$, $A_1=C_1=2$, (\ref{tps1}) becomes
\begin{eqnarray}
\begin{split}
p_1(t)&=\frac{1}{2}p_2(t)=e^{-\frac{3}{2}e^{-2}t}, \quad r_1(t)=2r_2(t)=2e^{\frac{3}{2}e^{-2}t},
\\
q_{1}(t)&=-\left(\frac{2}{3}+\frac{5}{2}e^{-2}\right)t+1, \qquad q_{2}(t)=-\left(\frac{2}{3}+\frac{5}{2}e^{-2}\right)t-1.
\end{split}
\label{f13pq}
\end{eqnarray}
and the associated solution of (\ref{eq}) with $b=0$ becomes
\begin{eqnarray}
\left\{\begin{array}{l}
u(x,t)=e^{-\frac{3}{2}e^{-2}t}\left(e^{- \left| x+(\frac{2}{3}+\frac{5}{2}e^{-2})t-1 \right|}
+2e^{- \left| x+(\frac{2}{3}+\frac{5}{2}e^{-2})t+1 \right|}\right),
\\
v(x,t)=e^{\frac{3}{2}e^{-2}t}\left(2e^{- \left| x+(\frac{2}{3}+\frac{5}{2}e^{-2})t-1 \right|}
+e^{- \left| x+(\frac{2}{3}+\frac{5}{2}e^{-2})t+1 \right|}\right).
\end{array} \right.
\label{tps13}
\end{eqnarray}
From (\ref{f13pq}), one can easily see that the amplitudes $p_1(t)$ and $p_2(t)$ of potential $u(x,t)$ are two monotonically decreasing functions of $t$,
while the amplitudes $r_1(t)$ and $r_2(t)$ of potential $v(x,t)$ are two monotonically increasing functions of $t$.
Figures \ref{f13u} and \ref{f13v} show the profiles of the potentials $u(x,t)$ and $v(x,t)$.

{\bf If $A_1\neq A_2$}, we may obtain the following solution of (\ref{tpode}):
\begin{eqnarray}
\begin{split}
p_1(t)&=Be^{\frac{3(A_2D^2-A_1)}{2D(A_1-A_2)}e^{- \frac{1}{3}\mid(A_1-A_2)t\mid}}, ~~
p_2(t)=\frac{p_1}{D},~~
r_1(t)=\frac{A_1}{p_1}, ~~
r_2(t)=\frac{A_2}{p_2},\\
q_{1}(t)&=-\frac{1}{3}A_1t+\frac{3(A_2D^2+A_1)}{2D(A_1-A_2)}sgn[(A_1-A_2)t]\left(e^{- \frac{1}{3}\mid(A_1-A_2)t\mid}-1\right),\\
q_{2}(t)&=-\frac{1}{3}A_2t+\frac{3(A_2D^2+A_1)}{2D(A_1-A_2)}sgn[(A_1-A_2)t]\left(e^{- \frac{1}{3}\mid(A_1-A_2)t\mid}-1\right),
\end{split}
\label{tps2}
\end{eqnarray}
where $B$ is an arbitrary integration constant. Let us study the following special cases of this solution.

\begin{figure}.
\begin{minipage}[t]{0.5\linewidth}
\centering
\includegraphics[width=2.2in]{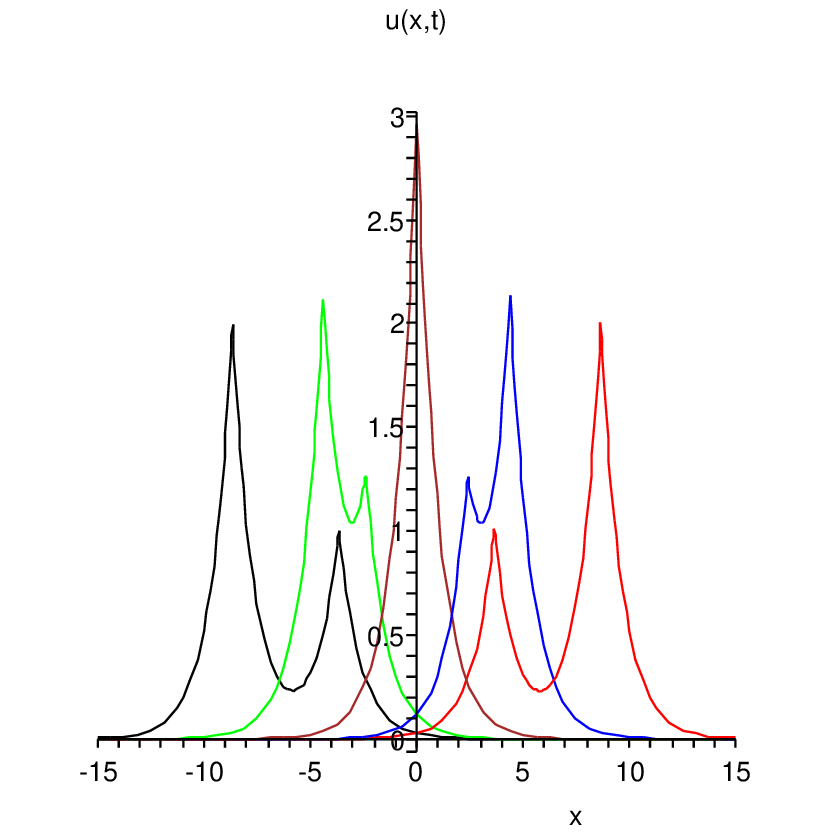}
\caption{\small{The two-peakon solution $u(x,t)$ in (\ref{tps21}). Red line: $t=-5$; Blue line: $t=-2$; Brown line: $t=0$ (collision); Green line: $t=2$; Black line: $t=5$.}}
\label{f21}
\end{minipage}%
\hspace{2.0ex}
\begin{minipage}[t]{0.5\linewidth}
\centering
\includegraphics[width=2.2in]{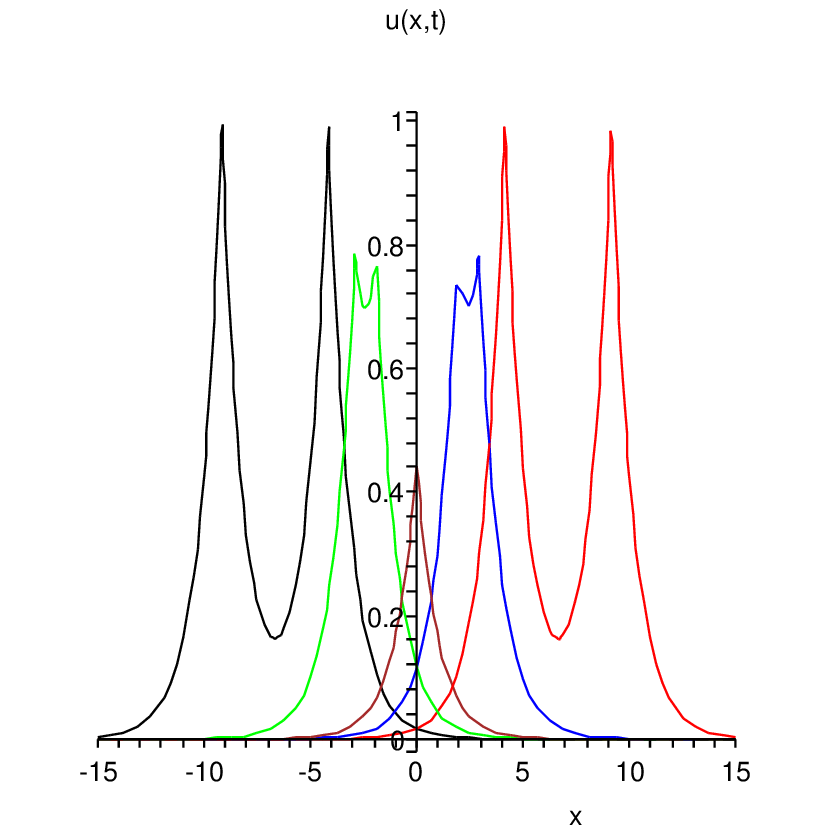}
\caption{\small{ The two-peakon solution $u(x,t)$ in (\ref{tps22}). Red line: $t=-5$; Blue line: $t=-1$; Brown line: $t=0$ (collision); Green line: $t=1$; Black line: $t=5$.}}
\label{f22u}
\end{minipage}
\end{figure}

{\sf Example 1}. Let $A_1=1$, $A_2=4$, $B=1$, $D=\frac{1}{2}$, then
\begin{eqnarray}
\left\{\begin{array}{l}
p_1(t)=r_1(t)=1, ~~p_2=r_2(t)=2,\\
q_{1}(t)=-\frac{1}{3}t+2sgn(t)\left(e^{- \mid t\mid}-1\right),\\
q_{2}(t)=-\frac{4}{3}t+2sgn(t)\left(e^{- \mid t\mid}-1\right).
\end{array}\right.
\label{case1}
\end{eqnarray}
The associated two-peakon solution of (\ref{eq}) becomes
\begin{eqnarray}
u(x,t)=v(x,t)=e^{-\left|x+\frac{1}{3}t-2sgn(t)\left(e^{- \mid t\mid}-1\right)\right|}+2e^{-\left|x+\frac{4}{3}t-2sgn(t)\left(e^{- \mid t\mid}-1\right)\right|}.
\label{tps21}
\end{eqnarray}
%$Q(t)\equiv q_{1}(t)-q_{2}(t)=t$.
As $t<0$ and $t$ is going to $0$, the tall peakon with the amplitude $2$ chases after the short peakon with the amplitude $1$. The two-peakon collides at time $t=0$.
After the collision ($t>0$), the peaks separate (the tall peakon surpasses the short one) and develop on their own way.
See Figure \ref{f21} for the detailed development of this kind of two-peakon.

\begin{figure}
\begin{minipage}[t]{0.5\linewidth}
\centering
\includegraphics[width=2.2in]{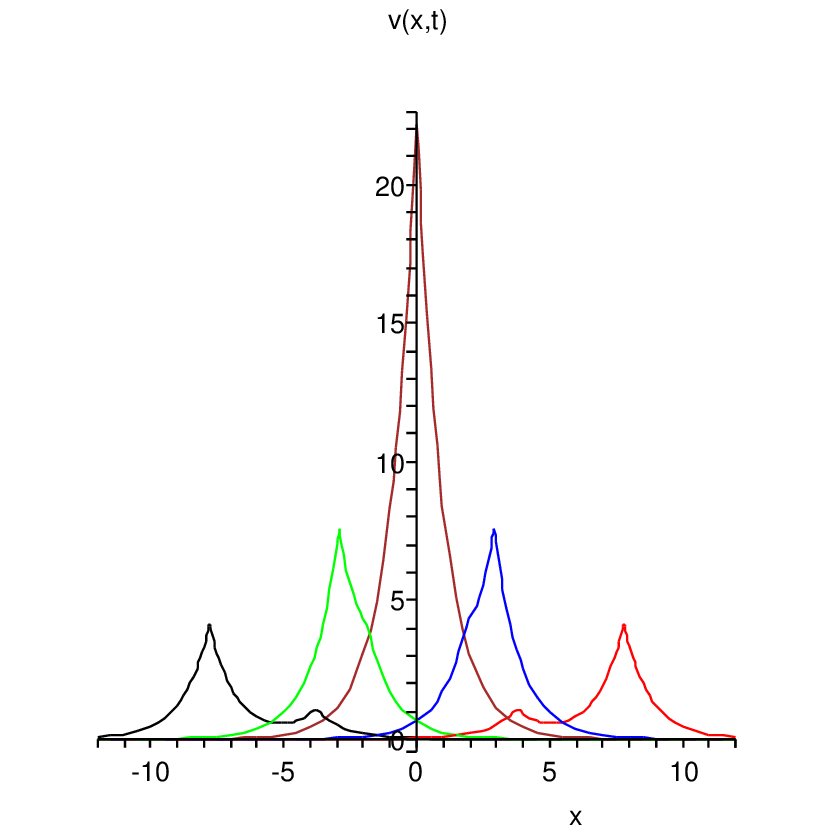}
\caption{\small{The two-peakon solution $v(x,t)$ in (\ref{tps22}).
Red line: $t=-4$; Blue line: $t=-1$; Brown line: $t=0$ (collision); Green line: $t=1$; Black line: $t=4$.}}
\label{f22v}
\end{minipage}
\hspace{2.0ex}
\begin{minipage}[t]{0.5\linewidth}
\centering
\includegraphics[width=2.2in]{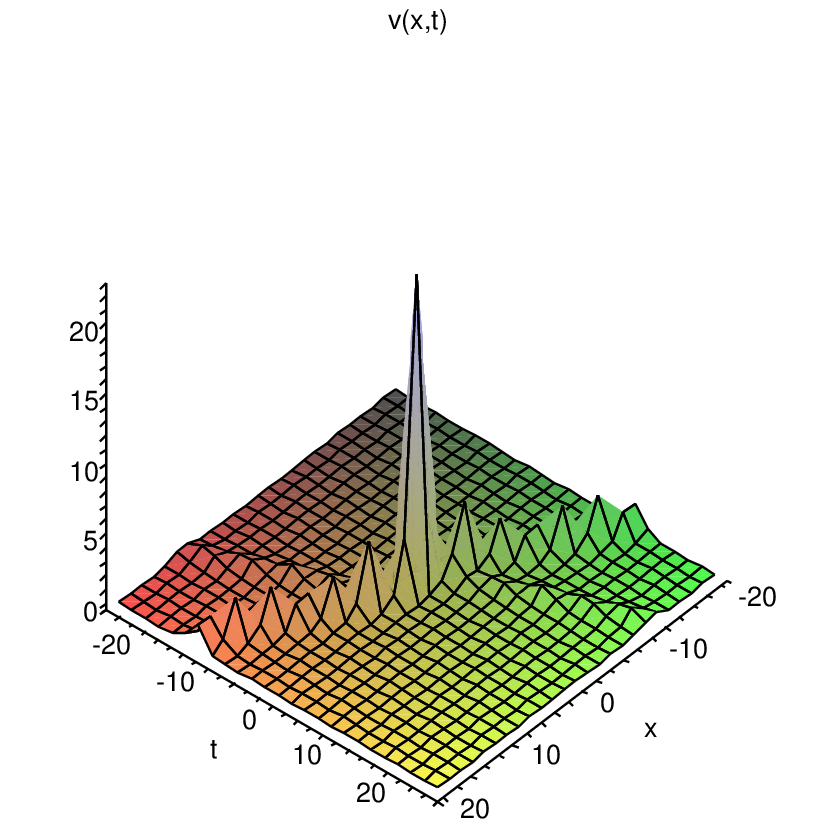}
\caption{\small{3-dimensional graph for the two-peakon solution $v(x,t)$ in (\ref{tps22}).}}
\label{f22v3d}
\end{minipage}
\end{figure}

{\sf Example 2}. Let $A_1=1$, $A_2=4$, $B=1$, $D=1$, then we have
\begin{eqnarray}
\left\{\begin{array}{l}
p_1(t)=p_2(t)=e^{-\frac{3}{2}e^{- \mid t\mid}}, \\
r_1(t)=e^{\frac{3}{2}e^{- \mid t\mid}},~~r_2(t)=4e^{\frac{3}{2}e^{- \mid t\mid}},\\
q_{1}(t)=-\frac{1}{3}t+\frac{5}{2}sgn(t)\left(e^{- \mid t\mid}-1\right),\\
q_{2}(t)=-\frac{4}{3}t+\frac{5}{2}sgn(t)\left(e^{- \mid t\mid}-1\right).
\end{array}\right.
\label{case2}
\end{eqnarray}
The associated two-peakon solution of (\ref{eq}) becomes
\begin{eqnarray}
\left\{\begin{array}{l}
u(x,t)=e^{-\frac{3}{2}e^{- \mid t\mid}}\left(e^{-\left|x+\frac{1}{3}t-\frac{5}{2}sgn(t)\left(e^{- \mid t\mid}-1\right)\right|}+e^{-\left|x+\frac{4}{3}t-\frac{5}{2}sgn(t)\left(e^{- \mid t\mid}-1\right)\right|}\right),\\
v(x,t)=e^{\frac{3}{2}e^{- \mid t\mid}}\left(e^{-\left|x+\frac{1}{3}t-\frac{5}{2}sgn(t)\left(e^{- \mid t\mid}-1\right)\right|}+4e^{-\left|x+\frac{4}{3}t-\frac{5}{2}sgn(t)\left(e^{- \mid t\mid}-1\right)\right|}\right).
\end{array}\right.
\label{tps22}
\end{eqnarray}
For the potential $u(x,t)$, the two-peakon solution possesses the same amplitude $e^{-\frac{3}{2}e^{- \mid t\mid}}$, which reaches the minimum value at $t=0$. Figure \ref{f22u} shows the profile of the two-peakon dynamics for the potential $u(x,t)$. For the potential $v(x,t)$, the two-peakon solution with the amplitudes $e^{\frac{3}{2}e^{- \mid t\mid}}$ and $4e^{\frac{3}{2}e^{- \mid t\mid}}$ collides at $t=0$. At this moment, the amplitudes attain the maximum value and the two-peakon overlaps into one peakon $5e^{\frac{3}{2}}e^{- \mid x \mid}$, which is much higher than other moments. See Figures \ref{f22v} and \ref{f22v3d} for the 2-dimensional and 3-dimensional graphs
of the two-peakon dynamics for the potential $v(x,t)$.

\begin{figure}
\begin{minipage}[t]{0.5\linewidth}
\centering
\includegraphics[width=2.2in]{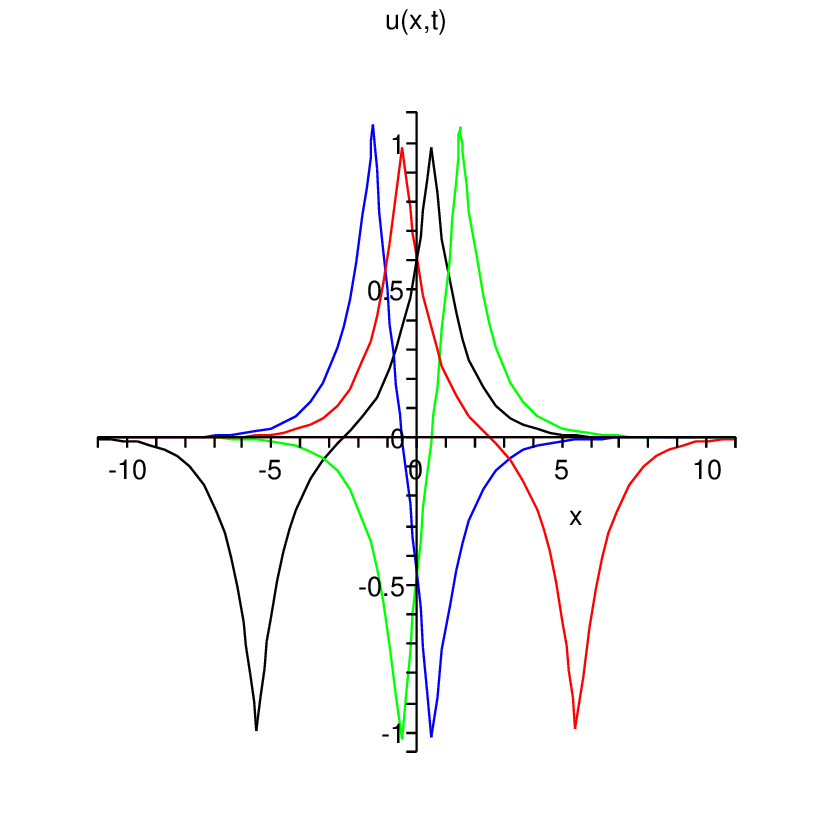}
\caption{\small{Peakon-antipeakon solution $u(x,t)$ in (\ref{tps23}). Red line: $t=-6$; Blue line: $t=-2$; At $t=0$ they collide and vanish; Green line: $t=2$; Black line: $t=6$.}}
\label{f23u}
\end{minipage}
\hspace{2.0ex}
\begin{minipage}[t]{0.5\linewidth}
\centering
\includegraphics[width=2.2in]{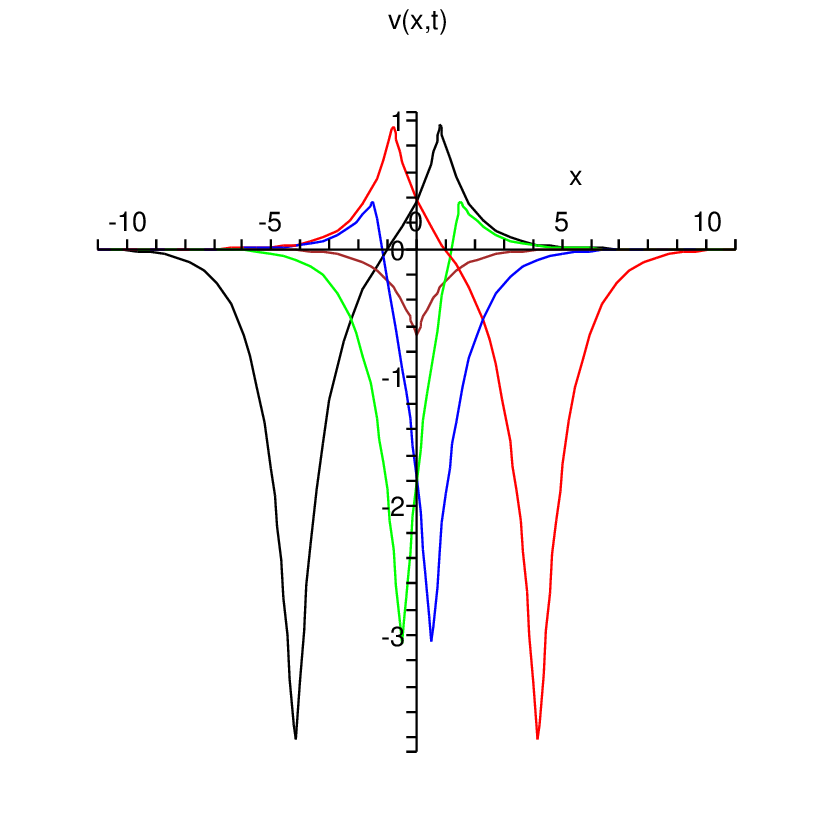}
\caption{\small{Peakon-antipeakon solution $v(x,t)$ in (\ref{tps23}). Red line: $t=-5$; Blue line: $t=-2$; Brown line: $t=0$ (collision); Green line: $t=2$; Black line: $t=5$.}}
\label{f23v}
\end{minipage}
\end{figure}

{\sf Example 3}. Let $A_1=1$, $A_2=4$, $B=1$, $D=-1$, then we have
\begin{eqnarray}
\left\{\begin{array}{l}
p_1(t)=-p_2(t)=e^{\frac{3}{2}e^{- \mid t\mid}}, \\
r_1(t)=e^{-\frac{3}{2}e^{- \mid t\mid}},~~r_2(t)=-4e^{-\frac{3}{2}e^{- \mid t\mid}},\\
q_{1}(t)=-\frac{1}{3}t-\frac{5}{2}sgn(t)\left(e^{- \mid t\mid}-1\right),\\
q_{2}(t)=-\frac{4}{3}t-\frac{5}{2}sgn(t)\left(e^{- \mid t\mid}-1\right).
\end{array}\right.
\label{case3}
\end{eqnarray}
The associated two-peakon solution of (\ref{eq}) becomes
\begin{eqnarray}
\left\{\begin{array}{l}
u(x,t)=e^{\frac{3}{2}e^{- \mid t\mid}}\left(e^{-\left|x+\frac{1}{3}t+\frac{5}{2}sgn(t)\left(e^{- \mid t\mid}-1\right)\right|}-e^{-\left|x+\frac{4}{3}t+\frac{5}{2}sgn(t)\left(e^{- \mid t\mid}-1\right)\right|}\right),\\
v(x,t)=e^{-\frac{3}{2}e^{- \mid t\mid}}\left(e^{-\left|x+\frac{1}{3}t+\frac{5}{2}sgn(t)\left(e^{- \mid t\mid}-1\right)\right|}-4e^{-\left|x+\frac{4}{3}t+\frac{5}{2}sgn(t)\left(e^{- \mid t\mid}-1\right)\right|}\right).
\end{array}\right.
\label{tps23}
\end{eqnarray}
For the potential $u(x,t)$, the peakon-antipeakon collides and vanishes at $t=0$. After the collision, the peakon and antipeakon reemerge and  separate. For the potential $v(x,t)$, the peakon and trough overlap at $t=0$, and then they separate. Figures \ref{f23u} and \ref{f23v} show the peakon-antipeakon dynamics for the potentials $u(x,t)$ and $v(x,t)$.

In general, we suppose an $N$-peakon solution has the following form
\begin{eqnarray}
u(x,t)=\sum_{j=1}^N p_j(t)e^{-\mid x-q_j(t)\mid}, ~~v(x,t)=\sum_{j=1}^N r_j(t)e^{-\mid x-q_j(t)\mid}.
\label{NP}
\end{eqnarray}
Substituting (\ref{NP}) into (\ref{eq}) with $b=0$ and integrating through test functions, we obtain the $N$-peakon dynamic system as follows
\begin{eqnarray}
\left\{\begin{array}{l}
p_{j,t}=\frac{1}{2}p_j\sum_{i,k=1}^N p_ir_k \left(sgn(q_j-q_k)-sgn(q_j-q_i)\right)e^{ -\mid q_j-q_k\mid-\mid q_j-q_i\mid},\\
q_{j,t}=\frac{1}{6}p_jr_j-\frac{1}{2}\sum_{i,k=1}^Np_ir_k\left(1-sgn(q_j-q_i)sgn(q_j-q_k)\right)e^{ -\mid q_j-q_i\mid-\mid q_j-q_k\mid},\\
r_{j,t}=-\frac{1}{2}r_j\sum_{i,k=1}^N p_ir_k \left(sgn(q_j-q_k)-sgn(q_j-q_i)\right)e^{ -\mid q_j-q_k\mid-\mid q_j-q_i\mid}.
\end{array}\right.   \label{dNcp}
\end{eqnarray}

\subsection{Peakon solutions to the two-component system (\ref{eq}) with different peakon positions}

In this section, we discuss the $N$-peakon %solutions of the two-component system in a more general case: peakon
solutions with different peakon positions based on the work in \cite{CHIP}. Let us assume that the $N$-peakon solutions of the two potentials $u$ and $v$ with different peakon positions are given in the form
\begin{eqnarray}
u(x,t)=\sum_{j=1}^N p_j(t)e^{-\mid x-q_j(t)\mid}, ~~v(x,t)=\sum_{j=1}^N r_j(t)e^{-\mid x-s_j(t)\mid},
\label{NPD}
\end{eqnarray}
where $q_i(t)\neq s_j(t),~1\leq i,j\leq N$.
With the help of delta functions, we have
\begin{eqnarray}
m(x,t)=2\sum_{j=1}^N p_j(t)\delta(x-q_j(t)), ~~n(x,t)=2\sum_{j=1}^N r_j(t)\delta(x-s_j(t)).
\label{NPD2}
\end{eqnarray}
Substituting (\ref{NPD}) and (\ref{NPD2}) into (\ref{eq}) with $b=0$ and integrating through test functions, we arrive at %obtain that $p_j$, $q_j$, $r_j$ and $s_j$ evolve according to
the following system regarding $p_j$, $q_j$, $r_j$, $s_j$:
\begin{eqnarray}
\left\{\begin{array}{l}
p_{j,t}=\frac{1}{2}p_j\sum_{i,k=1}^N p_ir_k \left(sgn(q_j-s_k)-sgn(q_j-q_i)\right)e^{ -\mid q_j-s_k\mid-\mid q_j-q_i\mid},\\
r_{j,t}=-\frac{1}{2}r_j\sum_{i,k=1}^N p_ir_k \left(sgn(s_j-s_k)-sgn(s_j-q_i)\right)e^{ -\mid s_j-s_k\mid-\mid s_j-q_i\mid},\\
q_{j,t}=-\frac{1}{2}\sum_{i,k=1}^Np_ir_k\left(1-sgn(q_j-q_i)sgn(q_j-s_k)\right)e^{ -\mid q_j-q_i\mid-\mid q_j-s_k\mid},\\
s_{j,t}=-\frac{1}{2}\sum_{i,k=1}^Np_ir_k\left(1-sgn(s_j-q_i)sgn(s_j-s_k)\right)e^{ -\mid s_j-q_i\mid-\mid s_j-s_k\mid}.
\end{array}\right.   \label{dNcpD}
\end{eqnarray}
Different from the $N$-peakon dynamic system of the coupled CH equation proposed in \cite{CHIP}, our system (\ref{dNcpD}) can not directly be rewritten in the standard form of a canonical Hamiltonian system. It is interesting to study whether (\ref{dNcpD}) is able to be rewritten as an integrable Hamiltonian system by introducing a Poisson bracket.
We will investigate the related topics in the near future.

\begin{figure}
\centering
\includegraphics[width=2.2in]{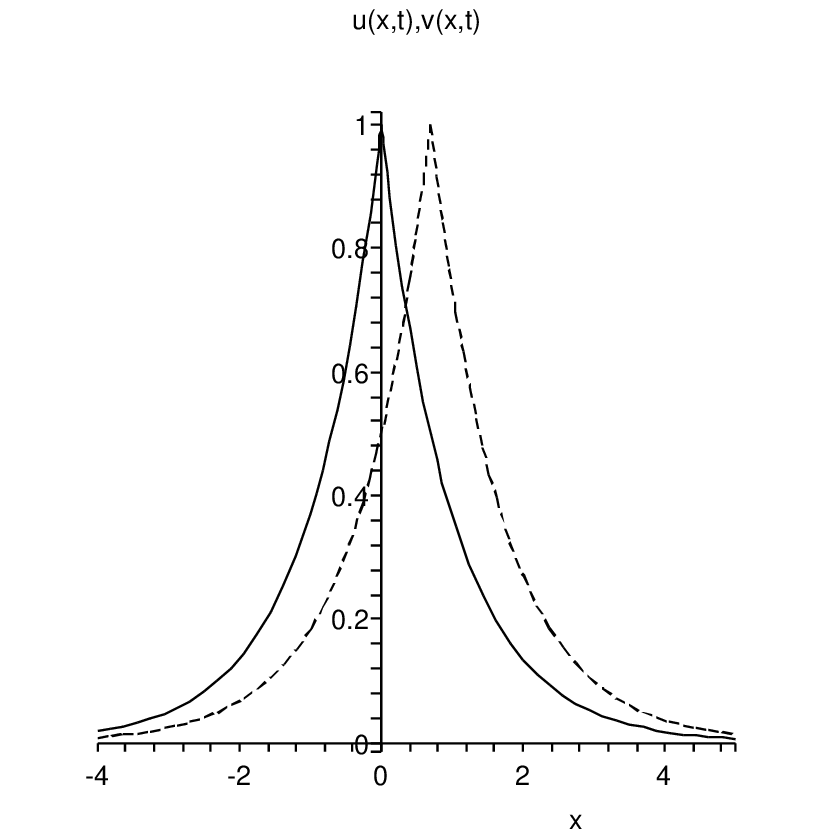}
\caption{\small{The single-peakon solution (\ref{Ds3}) at $t=0$. Solid line: $u(x,0)$; Dashed line: $v(x,0)$.}}
\label{fds}
\end{figure}

For $N=1$, (\ref{dNcpD}) is reduced to
\begin{eqnarray}
\left\{\begin{array}{l}
p_{1,t}=\frac{1}{2}p_1^{2}r_1 sgn(q_1-s_1)e^{ -\mid q_1-s_1\mid},\\
r_{1,t}=\frac{1}{2}p_1r_1^{2} sgn(s_1-q_1)e^{ -\mid s_1-q_1\mid},\\
q_{1,t}=-\frac{1}{2}p_1r_1e^{ -\mid q_1-s_1\mid},\\
s_{1,t}=-\frac{1}{2}p_1r_1e^{ -\mid s_1-q_1\mid}.
\end{array}\right.
\label{d1cpD}
\end{eqnarray}
From the last two equations of (\ref{d1cpD}), we obtain
\begin{eqnarray}
s_1=q_1+A_1,
\label{Ds1}
\end{eqnarray}
where $A_1\neq 0$ is an integration constant. Without loss of generality, we suppose $A_1>0$. Substituting (\ref{Ds1}) into (\ref{d1cpD}) leads to%, we finally arrive at
\begin{eqnarray}
\left\{\begin{array}{l}
p_{1}=A_3e^{-\frac{1}{2}e^{-A_1}A_2t},\\
r_{1}=\frac{A_2}{A_3}e^{\frac{1}{2}e^{-A_1}A_2t},\\
q_{1}=-\frac{1}{2}e^{-A_1}A_2t,\\
s_{1}=-\frac{1}{2}e^{-A_1}A_2t+A_1,
\end{array}\right.
\label{Ds2}
\end{eqnarray}
where $A_2$ and $A_3$ are integration constants.
In particular, we take $A_1=\ln2$ and $A_2=A_3=1$, then the single-peakon solutions with different peakon positions become
\begin{eqnarray}
u=e^{-\frac{1}{4}t}e^{-\mid x+\frac{1}{4}t\mid},\quad v=e^{\frac{1}{4}t}e^{-\mid x+\frac{1}{4}t-\ln2\mid}. \label{Ds3}
\end{eqnarray}
See Figure \ref{fds} for the profile of this single-peakon solution at $t=0$.
We have not yet explicitly solved (\ref{dNcpD}) with $N=2$. This is due to the complexity of (\ref{dNcpD}) with $N=2$, which is a coupled ordinary differential equation with eight components.

\section{Solutions to the integrable system (\ref{nlseq})}
As mentioned above, system (\ref{eq}) is cast into the integrable cubic nonlinear equation (\ref{nlseq})
under the complex conjugate reduction $v=u^\ast$. Thus equation (\ref{nlseq}) possesses the following Lax pair
\begin{eqnarray}
\left(\begin{array}{c}\phi_{1}\\\phi_{2} \end{array}\right)_x&=&
U\left(\begin{array}{c} \phi_{1}\\\phi_{2}
\end{array}\right),\quad
U=\frac{1}{2}\left( \begin{array}{cc} -\alpha & \lambda m\\
 -\lambda  m^\ast &  \alpha \\ \end{array} \right),
\label{LPS2}\\
\left(\begin{array}{c}\phi_{1}\\\phi_{2} \end{array}\right)_t&=&
V\left(\begin{array}{c} \phi_{1}\\\phi_{2}
\end{array}\right),\quad V=-\frac{1}{2}\left( \begin{array}{cc} V_{11} & V_{12} \\ V_{21} & -V_{11} \\ \end{array} \right),
\label{LPT2}
\end{eqnarray}
with $\alpha=\sqrt{1-\lambda^2b}$, and
\begin{eqnarray}
\begin{split}
V_{11}=& \lambda^{-2}\alpha+\frac{\alpha}{2}(|u|^2-|u_x|^2)+\frac{1}{2}(uu_x^*-u^*u_x),
\\ V_{12}=& -\lambda^{-1}(u-\alpha u_x)-\frac{1}{2}\lambda m(|u|^2-|u_x|^2),
\\ V_{21}=&\lambda^{-1}(v+\alpha v_x)+\frac{1}{2}\lambda n(|u|^2-|u_x|^2).
\end{split}
\label{ABC2}
\end{eqnarray}
Next, we show that the dispersionless version of equation (\ref{nlseq}) with $b=0$
admits the complex valued $N$-peakon solution, while the dispersion version of
equation (\ref{nlseq}) with $b\neq0$ allows the complex valued kink wave solution.

\subsection{Complex valued peakon solution of (\ref{nlseq}) with $b=0$}
Let us assume that a complex valued $N$-peakon solution of (\ref{nlseq}) with $b=0$ has the following form
\begin{eqnarray}
u=\sum_{j=1}^N \left(p_j(t)+\sqrt{-1} r_j(t)\right)e^{-\mid x-q_j(t)\mid},
\label{NP2}
\end{eqnarray}
where $p_j(t)$, $r_j(t)$ and $q_j(t)$ are real valued functions.
Substituting (\ref{NP2}) into (\ref{nlseq}) with $b=0$ and integrating through real valued test functions, and separating the real part and imaginary part,
we finally obtain that $p_j(t)$, $r_j(t)$ and $q_j(t)$ evolve according to the dynamical system
\begin{eqnarray}
\left\{
\begin{split}
p_{j,t}=&r_j\sum_{l,k=1}^N p_lr_k \left(sgn(q_j-q_k)-sgn(q_j-q_l)\right)e^{ -\mid q_j-q_k\mid-\mid q_j-q_l\mid},\\
r_{j,t}=&p_j\sum_{l,k=1}^N p_lr_k \left(sgn(q_j-q_l)-sgn(q_j-q_k)\right)e^{ -\mid q_j-q_k\mid-\mid q_j-q_l\mid},\\
q_{j,t}=&\frac{1}{6}(p_j^2+r_j^2)-\frac{1}{2}\sum_{l,k=1}^N(p_lp_k+r_lr_k)\left(1-sgn(q_j-q_l)sgn(q_j-q_k)\right)e^{ -\mid q_j-q_l\mid-\mid q_j-q_k\mid}.
\end{split}
\right.\label{dNcp2}
\end{eqnarray}

For $N=1$, (\ref{dNcp2}) becomes
\begin{eqnarray}
p_{1,t}=0,\quad
r_{1,t}=0,\quad
q_{1,t}=-\frac{1}{3}(p_1^2+r_1^2),
\label{dNcp21}
\end{eqnarray}
which gives
\begin{eqnarray}
p_{1}=c_1,\quad
r_{1}=c_2,\quad
q_{1}=-\frac{1}{3}(c_1^2+c_2^2)t,
\label{dNcp21s}
\end{eqnarray}
where $c_{1}$ and $c_{2}$ are real valued integration constants. Thus we arrive at the single-peakon solution
\begin{eqnarray}
u=(c_1+\sqrt{-1} c_{2})e^{-\mid x+\frac{c_1^2+c_2^2}{3}t\mid}=ce^{-\mid x+\frac{1}{3}|c|^2t\mid},\label{ocpnp2}
\end{eqnarray}
where $c=c_{1}+\sqrt{-1} c_{2}$, and $|c|$ is the modulus of $c$.

For $N=2$,
%(\ref{dNcp2}) becomes
%\begin{eqnarray}
%\left\{\begin{split}
%q_{1,t}=&-\frac{1}{3}(p_1^2+r_1^2)- (p_1p_2+r_1r_2)e^{ -\mid q_1-q_2\mid},\\
%q_{2,t}=&-\frac{1}{3}(p_2^2+r_2^2)- (p_1p_2+r_1r_2)e^{ -\mid q_1-q_2\mid},\\
%p_{1,t}=&r_1(p_1r_2-p_2r_1)sgn(q_1-q_2)e^{ -\mid q_1-q_2\mid},\\
%p_{2,t}=&r_2(p_1r_2-p_2r_1)sgn(q_1-q_2)e^{ -\mid q_1-q_2\mid},\\
%r_{1,t}=&-p_1(p_1r_2-p_2r_1)sgn(q_1-q_2)e^{ -\mid q_1-q_2\mid},\\
%p_{2,t}=&-p_2(p_1r_2-p_2r_1)sgn(q_1-q_2)e^{ -\mid q_1-q_2\mid}.
%\end{split}\right.\label{dtcpnp}
%\end{eqnarray}
we may solve (\ref{dNcp2}) as
\begin{eqnarray}
\left\{\begin{array}{l}
q_{1}(t)=-\frac{1}{3}A_1^2t+\Gamma_{1}(t),\\
q_{2}(t)=-\frac{1}{3}A_2^2t+\Gamma_{1}(t),\\
p_{1}(t)=A_1\sin(\Gamma_{2}(t)+A_3),\\
p_{2}(t)=A_2\sin(\Gamma_{2}(t)+A_4),\\
r_{1}(t)=A_1\cos(\Gamma_{2}(t)+A_3),\\
r_{2}(t)=A_2\cos(\Gamma_{2}(t)+A_4),
\end{array}\right. \label{2pq}
\end{eqnarray}
where
\begin{eqnarray}
\begin{split}
\Gamma_{1}(t)&=\frac{3A_1A_2\cos(A_3-A_4)}{|A_1^2-A_2^2|}sgn(t)\left(e^{-\frac{1}{3}\mid(A_1^2-A_2^2) t\mid}-1\right),\\ \Gamma_{2}(t)&=\frac{3A_1A_2\sin(A_3-A_4)}{A_1^2-A_2^2}e^{-\frac{1}{3}\mid(A_1^2-A_2^2) t\mid},
\end{split}
\label{gma}
\end{eqnarray}
and $A_1$, $\cdots$, $A_4$ are real valued integration constants.
Hence the two-peakon solution reads
\begin{eqnarray}
u=A_1\sqrt{-1} e^{-\sqrt{-1} (\Gamma_{2}(t)+A_3)}e^{-\mid x+\frac{1}{3}A_1^2t-\Gamma_{1}(t)\mid}+A_2\sqrt{-1} e^{-\sqrt{-1}(\Gamma_{2}(t)+A_4)}e^{-\mid x+\frac{1}{3}A_2^2t-\Gamma_{1}(t)\mid},
\label{2su}
\end{eqnarray}
where the Euler formula $e^{\sqrt{-1} x}=cosx+\sqrt{-1} sinx$ is employed.

\subsection{Complex valued kink solution of (\ref{nlseq}) with $b\neq0$}

We suppose that a complex valued kink wave solution of equation (\ref{nlseq}) with $b\neq0$ has the form
\begin{eqnarray}
u=\left(C_1+\sqrt{-1} C_2\right)sgn(x-ct)\left(e^{-\mid x-ct\mid}-1\right),
\label{kink12}
\end{eqnarray}
where the real constant $c$ is the wave speed, and $C_1$ and $C_2$ are two real constants to be determined.
Substituting (\ref{kink12}) into equation (\ref{nlseq}) with $b\neq0$ and integrating through real valued test functions,
and separating its real part and imaginary part,
we finally arrive at
\begin{eqnarray}
c=-\frac{1}{2}b,~~
C_1^2+C_2^2=-b.
\label{ikss}
\end{eqnarray}

Formula (\ref{ikss}) implies that the wave speed is exactly $-\frac{1}{2}b$, where $b<0$ is the coefficient of the linear dispersive term.
Thus the  complex valued weak kink solution becomes
\begin{eqnarray}
u=Csgn(x+\frac{1}{2}b t)\left(e^{-\mid x+\frac{1}{2}bt\mid}-1\right),
\label{kink12s}
\end{eqnarray}
where $C=C_1+\sqrt{-1} C_2$, and $|C|^2=-b$, $b<0$. We remark that in (\ref{kink12s}) only the constant $C$ is complex, the variables $x$ and $t$ are real valued variables.

%It is very interesting and important to study whether equation (\ref{nlseq}) possesses the soliton solutions like the form of the soliton solution of the  standard nonlinear Schr\"{o}dinger equation. We will make a further study for these topics.

\section {Conclusions and discussions}

In our paper, we propose a new integrable two-component system with cubic nonlinearity and linear dispersion. The system is shown to possess Lax pair, bi-Hamiltonian structure, and infinitely many conservation laws. Geometrically, this system describes a nontrivial one-parameter family of pseudo-spherical surfaces. In the dispersionless case, we show the system admits $N$-peakon solution and explicitly solve the system for the single-peakon and the two-peakon dynamical system. Moreover, we propose a scalar integrable complex cubic nonlinear equation and find the complex valued $N$-peakon solution and kink wave solution to the integrable complex equation.

In \cite{SQQ}, the authors introduced an integrable two-component extension of the dispersionless version of cubic nonlinear equation (\ref{cCHQ}) (or the FORQ equation called in some literature)
\begin{eqnarray}
\left\{\begin{array}{l}
m_t=-[m(uv-u_xv_x+u_xv-uv_x)]_x,
\\ n_t=-[ n(uv-u_xv_x+u_xv-uv_x)]_x,
\\m=u-u_{xx},~~ n=v-v_{xx}.
\end{array}\right. \label{SQQ}
\end{eqnarray}
We remark that the dispersionless version of our two-component system (\ref{eq}) with $b=0$ is not equivalent to system (\ref{SQQ}). System (\ref{eq}) in our paper is able to be reduced to the CH equation, but system (\ref{SQQ}) is not, which apparently implies that these two equations are not equivalent. In fact, both system (\ref{eq}) with $b=0$ and system (\ref{SQQ}) belong to a more general negative flow in a hierarchy.
For the details of this topic, one may see our very recent paper \cite{XQZ}.

It is an interesting task to study whether there are other new features in the structure of solutions for our two-component system, and particularly for our  complex equation with a linear dispersive term. Also other topics, such as smooth soliton solutions \cite{MY1}, cuspons, peakon stability, and algebra-geometric solutions, remain open for our system (\ref{eq}) and (\ref{nlseq}).

\vspace*{0.5cm}
\noindent {\bf Data accessibility.}
There are no primary data in this article.

\noindent {\bf Competing interests.}
There are no competing interests.

\noindent {\bf Authors' contributions.}
This is a theoretical paper. The author Xia is the first author. The author Qiao is the corresponding author.

\noindent {\bf Acknowledgements.}
The authors would like to express their sincerest thanks to the reviewers for their valuable
comments and suggestions. The author Xia also thanks Ruguang Zhou for helpful discussions.

\noindent {\bf Funding statement.}
The author Xia was supported by the National Natural Science Foundation of China (Grant Nos. 11301229 and 11271168), the Natural Science Foundation of the Jiangsu Province (Grant No. BK20130224)
and the Natural Science Foundation of the Jiangsu Higher Education Institutions of China (Grant No. 13KJB110009).
The author Qiao was partially supported by the National Natural Science Foundation of China (Grant No. 11171295 and 61328103) and also thanks
the U.S. Department of Education GAANN project (P200A120256) to support UTPA mathematics graduate program.

\vspace{1cm}
\small{

}
\end{document}